\documentclass[lettersize,journal]{IEEEtran}
\usepackage{amsmath,amsfonts}
\usepackage{amsthm}
\usepackage{algorithmic}
\usepackage{array}
\usepackage{textcomp}
\usepackage{stfloats}
\usepackage{url}
\usepackage{cite}
\usepackage{verbatim}
\usepackage{graphicx}
\usepackage{xcolor}
\def\BibTeX{{\rm B\kern-.05em{\sc i\kern-.025em b}\kern-.08em
    T\kern-.1667em\lower.7ex\hbox{E}\kern-.125emX}}
\usepackage{balance}
\usepackage[font=scriptsize]{caption}
\usepackage{subcaption}

\usepackage{subfig}

\usepackage{algorithmic}
\usepackage{algorithm}

\newcommand{\calG}{{\mathcal{G}}}

\newcommand{\R}{\mathbb{R}}
\newcommand{\N}{\mathcal{N}}

\newtheorem{theorem}{Theorem}[section]
\newtheorem{lemma}[theorem]{Lemma}

\newtheorem{corollary}[theorem]{Corollary}

\newtheorem{problem}{Problem}
\newtheorem{remark}[theorem]{Remark}
\newtheorem{assumption}{Assumption}

\newcommand{\real}{\mathbb{R}}

\newcommand{\complex}{\mathbb{C}}

\newcommand{\veci}{\boldsymbol{i}}
\newcommand{\vecv}{\boldsymbol{v}}
\newcommand{\vecV}{\boldsymbol{\nu}}
\newcommand{\vecs}{\boldsymbol{s}}
\newcommand{\vecP}{\boldsymbol{p}_{\text{net}}}
\newcommand{\vecQ}{\boldsymbol{q}_{\text{net}}}
\newcommand{\vecp}{\boldsymbol{p}}
\newcommand{\vecq}{\boldsymbol{q}}
\newcommand{\vecpl}{\boldsymbol{p}_l}
\newcommand{\vecql}{\boldsymbol{q}_l}
\newcommand{\smally}{\boldsymbol{\bar{y}}}
\newcommand{\capitalY}{\boldsymbol{Y}}
\newcommand{\diag}{\text{diag}}
\newcommand{\vecu}{\boldsymbol{u}}

\DeclareMathAlphabet{\mymathbb}{U}{BOONDOX-ds}{m}{n}

\newcommand{\zero}{\mymathbb{0}}

\allowdisplaybreaks[3]

\begin{document}
\title{Optimal Power Flow Pursuit \\ via Feedback-based Safe Gradient Flow}
\author{Antonin Colot$^*$, Yiting Chen$^*$, Bertrand Corn\'{e}lusse, Jorge Cort\'{e}s, and Emiliano Dall'Anese 
\thanks{$^*$ Equal contribution of the authors.}
\thanks{This work was supported in part by the
National Science Foundation (NSF) awards 1941896 and 1947050. The work of Antonin Colot was supported by the Research Fellow Fellowship of the F.R.S-FNRS.}
\thanks{\emph{(Corresponding author: Antonin Colot.)}}
\thanks{Antonin Colot is with the Montefiore Institute, University of Li\`{e}ge, B-4000 Liège, Belgium, and a visiting student at the University of Colorado Boulder, CO 80309 Boulder, USA (email: antonin.colot@uliege.be).}
\thanks{Yiting Chen is with the Department of Electrical and Computer Engineering, Boston University, Boston, MA 02215, USA (email: yich4684@bu.edu).}
\thanks{Bertrand Corn\'{e}lusse is with the Montefiore Institute, University of Liège, B-4000 Liège, Belgium (email: bertrand.cornelusse@uliege.be).}
\thanks{Jorge Cort\'{e}s is with the Department of Mechanical and Aerospace Engineering, University of California San Diego, CA 92093 San Diego, USA (email: cortes@ucsd.edu).}
\thanks{Emilliano Dall'Anese is with the Department of Electrical and Computer Engineering and the Division of Systems Engineering, Boston University, Boston, MA 02215, USA (email: edallane@bu.edu).}
}

\maketitle

\begin{abstract}
This paper considers the problem of controlling inverter-interfaced distributed energy resources (DERs) in a distribution grid to solve an AC optimal power flow (OPF) problem in real time. The AC OPF includes voltage constraints, and seeks to minimize costs associated with the economic operation, power losses, or the power curtailment from renewables.  We develop an online feedback optimization method to drive the DERs' power setpoints to solutions of an AC OPF problem based only on voltage measurements (and without requiring measurements of the power consumption of non-controllable assets). The proposed method -- grounded on the theory of control barrier functions -- is based on a continuous approximation of the projected gradient flow, appropriately modified to accommodate measurements from the power network. We provide results in terms of local exponential stability, and assess the robustness to errors in the measurements and in the system Jacobian matrix. We show that the proposed method ensures anytime satisfaction of the voltage constraints when no model and measurement errors are present; if these errors are present and are small,
the voltage violation is practically negligible. We also discuss extensions of the framework to virtual power plant setups and to cases where constraints on power flows and currents must be enforced. Numerical experiments on a 93-bus distribution system and with realistic load and production profiles show a superior performance in terms of voltage regulation relative to existing methods. 
\end{abstract}


\section{Introduction}

This work seeks to contribute to the domain of real-time control and operation of distribution systems with high integration of inverter-interfaced distributed energy resources (DERs). The steady increase in energy costs, combined with government incentives advocating for the utilization of renewable energy sources and of energy-efficient automated load control, has reshaped the operation of distribution networks~\cite{taylor2016power,kroposki2020autonomous}. Historically, distribution networks were designed to manage unidirectional power flows; however, the increased integration of renewable resources and load management strategies lead to  operational and reliability challenges related to reversed power flows, voltage fluctuations, and power quality. 

Traditional techniques based on solving an AC optimal power flow (OPF) problem~\cite{zhu2007optimal,capitanescu2016critical} require collecting information of all non-controllable powers and
running an iterative method; this process may be long compared to the fast changing conditions of a modern distribution
system~\cite{taylor2016power,kroposki2020autonomous}; existing Volt/Var techniques may not fully resolve voltage regulation and may in fact increase line currents; recent works on emulating OPF solutions via neural networks can alleviate the computational burden~\cite{baker2020emulating,nellikkath2022physics}, but still require measurements of all the non-controllable powers (which are the inputs to the neural network), and may not even produce feasible power setpoints. 

%

Collecting measurements of non-controllable powers in distribution networks in real time (e.g., at the second level) is challenging as distribution systems are historically measurement-scarce, as discussed in~\cite{cheng2023survey}. Even though new types of sensing and communication infrastructures can capture real-time data, 
%
%
traditional techniques for solving the AC-OPF problem in distribution systems require collecting load measurements in real time from \emph{each} meter and distribution transformer, which is impractical and economically unfeasible~\cite{muscas2014effects}.
In this work, we focus on real-time AC OPF methods~\cite{bolognani2014distributed,dall2016optimal,Hauswirth16,Hauswirth17,nowak2020measurement,ortmann2023deployment}, and seek new strategies that exhibit strong performance in terms of achieved operational cost and voltage limit satisfaction (both from analytical and numerical standpoints), while using limited  measurements. In particular, we seek methods that do not require a complete AC model and knowledge of all the non-controllable powers throughout the nodes of the system. 

\emph{Prior work}. Several approaches have been explored to develop real-time OPF algorithms. 
In general, existing  solutions leverage online optimization techniques,  and incorporate  measurements of some network quantities to bypass the need of a system-level model. In the following, we present a list that is by no means exhaustive. Feedback algorithms using voltage measurements based on linearized models were developed in~\cite{bolognani2014distributed}, and recently~\cite{ZY-GC-MKS-JC:23-tps} combined with data-driven learning to synthesize decentralized strategies; online primal-dual methods with voltage and/or power measurements have been proposed in \cite{dall2016optimal, bernstein2019real}; model-free counterparts were proposed in \cite{chen2020model} and \cite{olives2022model}. Discrete-time projected gradient algorithms for the OPF problem are employed in \cite{gan2016online,nowak2020measurement}, while projected gradient flows were used in \cite{Hauswirth16}. Projection of gradient iterates onto a linearization of the feasible set around the current state was used in~\cite{ortmann2023deployment} for reactive power control. Power control for aggregations of DERs to track setpoints at the point of common coupling via gradient-type methods were proposed in, e.g.,~\cite{bernstein2015composable}. Online quasi-Newton methods were used in \cite{Tang17}, and online interior-point methods were proposed in~\cite{lupien2023online}. In addition to \cite{bolognani2014distributed}, distributed methods were explored in, e.g.,~\cite{colot2023fully}.

\emph{Contributions}. Compared to the works in the context of real-time OPF methods mentioned above, the contributions of our paper can be described as follows:

\noindent \emph{(c1)} We propose a new approach for 
the design of real-time OPF algorithms that is  grounded on the theory of control barrier functions (CBFs)~\cite{ames2019control}. We leverage a continuous approximation of projected gradient flows~\cite{allibhoy2023control}, appropriately modified to accommodate voltage measurements from the power network. Inheriting the properties of CBF methods, the proposed algorithm -- here termed feedback-based safe gradient flow (SGF) -- ensures anytime satisfaction of
the voltage constraints, while reaching solutions of the OPF.  

\noindent \emph{(c2)} From a theoretical standpoint, we show that the proposed  feedback-based SGF renders 
isolated optimal solutions of the AC OPF problem locally exponentially stable and ensures the anytime satisfaction of the voltage constraints. On the other hand, existing feedback-based optimization methods for distribution systems~\cite{bolognani2014distributed,dall2016optimal,Hauswirth16,Hauswirth17,nowak2020measurement,ortmann2023deployment} do not guarantee anytime satisfaction of
the voltage constraints.

\noindent \emph{(c3)} We provide results in terms of practical exponential stability and practical forward invariance when voltage measurements are affected by errors and when the Jacobian matrix of the AC power flow equations is computed only approximately (for example, when a linear approximation of the power flow equations is used).

\noindent \emph{(c4)} We perform numerical experiments on a 93-bus distribution system \cite{SimBench} and with realistic load and solar production
profiles from the Open Power System 
Data. We show that our method shows far superior performance in terms of voltage
regulation relative to existing online primal-dual methods and Volt/Var strategies.

We note that, relative to~\cite{ortmann2023deployment}, our design leverages the theory of CBFs~\cite{ames2019control,allibhoy2023control}, our method can handle constraints that are nonlinear, and we provide  practical stability and forward-invariance guarantees. 

From a practical standpoint, we provide remarks throughout the paper on how the proposed method can be integrated into the existing distribution system infrastructure: for instance, by leveraging distributed energy resources management systems (DERMSs) for distribution operators and existing communication infrastructure or a Supervisory Control and Data Acquisition (SCADA) system.



\section{Problem formulation}
\label{sec:problemformulation}

\emph{Notation}. Upper-case (lower-case) boldface letters are used for matrices (column vectors); $(.)^\top$ denotes the transposition and $(.)^*$ the complex-conjugate; 
$j$ the imaginary unit and $|.|$ the absolute value of a number. If we consider a given vector $\boldsymbol{x} \in \real^{N}$, $\text{diag}(\cdot)$ returns a $N\times N$ matrix with the element of $\boldsymbol{x}$ in its diagonal. For vectors \( \boldsymbol{x} \in \R^n \) and \( \boldsymbol{u} \in \R^m \),
\( \| \boldsymbol{x} \| \) denotes the $\ell_2$-norm and  \( (\boldsymbol{x},\boldsymbol{u}) \in \R^{n + m}\) denotes their vector concatenation. We denote as $\zero$ a vector or matrix with all zeros (the dimensions will be clear from the context). $\mathbb{C}$ denotes the set of complex numbers and for a vector $\boldsymbol{x} \in \mathbb{C}^{N}$, $\Re(\boldsymbol{x}) \in \mathbb{R}^{N}$ denotes its real part and $\Im(\boldsymbol{x}) \in \mathbb{R}^{N}$ its imaginary part.

\subsection{Distribution system model}

Consider an electrical distribution system with $N+1$ nodes
%
%
and hosting $G$ distributed energy resources (DERs); these may include inverter-interfaced photovoltaic systems, energy storage systems, variable-speed drives, and  
electric vehicles, or small-scale generators if any.  The node $0$ is taken to be the substation or the point of common coupling, while $\N := \{1,...,N\}$ is the set of remaining  nodes. We consider a steady-state model where voltages and currents are represented in the phasor domain. Accordingly, let  $v_k = \nu_k e^{j\delta_k} \in \complex$, $\nu_k := |v_k|$,
and  $i_k = |i_k| e^{j\psi_k} \in \complex$ the line-to-ground voltage and current injected at node $i$, respectively. Moreover, the voltage at  node $0$ is set to $v_0 = V_0 e^{j\delta_0}$~\cite{Kerstingbook}.

Using Ohm's Law and Kirchhoff's Law, one has the usual phasor relationship: 
\begin{equation}
\label{eq:Ymatrix}
    \begin{bmatrix}
        i_0\\
        \veci
    \end{bmatrix}
     = 
    \begin{bmatrix}
        y_0 & \smally^T\\
        \smally & \capitalY\\
    \end{bmatrix}  
    \begin{bmatrix}
        v_0\\
        \vecv
    \end{bmatrix}
\end{equation}
where $\vecv \in \complex^{N}$ collects the voltages $\{v_k\}_{k \in \N}$, $\veci \in \complex^{N}$ collects the currents $\{i_k\}_{k \in \N}$, and $\capitalY \in \complex^{N\times N}$, $\smally \in \complex^{N}$, and $y_0 \in \complex$ are based on the series and shunt admittances of the distribution lines represented by a standard $\Pi$-model; see, for example,~\cite{Kerstingbook,bolognani2015existence}. Using \eqref{eq:Ymatrix}, it is possible to relate complex powers at the nodes $\N$ with voltages as
\begin{equation}\label{eq:pfequations}
    \vecs = \diag(\vecv)\left(\smally^*v_0^* + \capitalY^*\vecv^*\right)
\end{equation}
where $\vecs = \vecP+\text{j}\vecQ \in \complex^{N}$, with $\vecP = [p_{\mathrm{net},1},...,p_{\mathrm{net},N}]^\top$ and $\vecQ = [q_{\mathrm{net},1},...,q_{\mathrm{net},N}]^\top$ vectors collecting the net active and reactive power injections at nodes $\mathcal{N}$. Note that $\vecP$ and $\vecQ$ account for both the powers (injected or consumed) of the DERs and the aggregate powers of the non-controllable loads that are connected to each of the nodes~$\N$. In particular, let $\vecpl := [p_{l,1}, \dots, p_{l,N}]^\top \in \mathcal{W}_p, \vecql := [q_{l,1}, \dots, q_{l,N}]^\top \in \mathcal{W}_q$ with a compact sets $\mathcal{W}_p\subset \mathbb{R}^{N}$ and $\mathcal{W}_q\subset \mathbb{R}^{N}$, be vectors collecting the net active and reactive power consumed at the nodes by non-controllable devices (positive when the power is consumed). For the $G$ DERs, consider the vector $\vecu = [p_1,p_2,...,p_G, q_1, q_2, ..., q_G]^\top$ collecting their active and reactive powers (with a positive sign denoting generation). Moreover, let $\calG := \{1,...,G\}$ be the index for the DERs, and define a function $m: \calG \rightarrow \mathcal{N}$ which maps a DER index to the node where the DER is connected to. With this notation, note that $\calG_n := \{i \in \calG: n = m(i)\}$ is the set of DERs connected at node~$n \in \N$. Then, the net active and reactive powers are given by $p_{\text{net},n} = \sum_{j \in \calG_n} p_j - p_{l,n}$ and $q_{\text{net},n} = \sum_{j \in \calG_n} q_j - q_{l,n}$ at each~$n$. In what follows, we consider a set of nodes $\mathcal{M} \subseteq \mathcal{N}$ with cardinality $M = |\mathcal{M}|$ where voltages are to be regulated (if the operator would like to monitor and regulate all the voltages, then $\mathcal{M} = \mathcal{N}$).
Finally, let $\mathcal{U}_i \subset \mathbb{R}^2$ be a compact set of admissible power setpoints, such that $(p_i,q_i) \in \mathcal{U}_i$ for $i\in\mathcal{G}$. We define $\mathcal{U} := \mathcal{U}_1 \times \mathcal{U}_2 \times \ldots \times \mathcal{U}_G$ so that $\vecu\in\mathcal{U}$.

Equation~\eqref{eq:pfequations} describes the power flow equations. For a given vector of net power injection $\vecs$, one can solve this nonlinear system of equations using numerical methods to find the vector of voltage phasors $\vecv$. Notice that the system of equations~\eqref{eq:pfequations} may have no, one, or many solutions. For the rest of this paper, we make the following assumption when restricting to a neighborhood of the nominal voltage profile.
\begin{assumption}[\textit{Mapping in  a neighborhood of the nominal voltage profile}]
\label{as:steadyStateMap}
There exists a unique continuously differentiable function $H:\mathcal{U} \times \mathcal{W}_p \times  \mathcal{W}_q \rightarrow  \mathbb{R}^M$  
such that, $H_i(\vecu; \vecpl, \vecql)= \nu_i = |v_i|$, for $i \in \mathcal{M}$. The Jacobian $J_H(\vecu;\vecpl,\vecql):=\frac{\partial H(\vecu; \vecpl, \vecql)}{\partial \vecu}$ is locally Lipschitz continuous.  \hfill $\Box$
\end{assumption}
If multiple solutions exist, we only consider the \emph{practical} solution, i.e., in the neighborhood of the nominal voltage profile, we restrict the attention to the solution that leads to high voltages and small line currents. The existence of the map $H$ is based on the Implicit Function Theorem and the results of, e.g.,~\cite{bolognani2015existence,bernstein2018load,wang2017existence} for single-phase and multi-phase distribution networks.  
\begin{remark}[\textit{Jacobian of map $H$}]
{\rm
For the sake of generality, we write the Jacobian   $J_H(\vecu;\vecpl,\vecql)$ as  dependent on the controllable and non-controllable power injections. However, in the paper we will leverage approximations of the Jacobian matrix; these estimates can be obtained without any knowledge of non-controllable power injections. We will provide remarks on linear approximations in Section~\ref{sec:linear}. \hfill $\Box$
}
\end{remark}

\begin{remark}[\textit{Model and notation}]
\label{remark: Model}
{\rm
It is important to note that the framework proposed in this paper works for multi-phase distribution systems with both wye and delta connections under the same Assumption~\ref{as:steadyStateMap}. The existence of the map $H$ for unbalanced multi-phase networks is discussed in~\cite{wang2017existence}. However, to simplify the notation and streamline the exposition, we outline the framework using a single-phase model. \hfill $\Box$ 
}
\end{remark}

\subsection{OPF for voltage regulation in distribution systems}
\label{sec:probformulation}


In this section, we outline a formulation of the OPF problem for distribution systems. By solving an OPF problem, one seeks power setpoints for the DERs that minimize the operational cost (or maximize performance objectives)   for the utilities and the customers, subject to operational constraints that may include voltage limits, line ampacity, or hardware limits~\cite{zhu2007optimal, capitanescu2016critical}. The cost associated with the utility companies may favor the minimization of system losses or the usage of controllable resources, or may perform voltage regulation (e.g., thus including cost of active power curtailment or reactive power compensation~\cite{bolognani2014distributed,colot2023fully});  on the other hand, customers may want to minimize the power curtailed by renewables or maximize their revenue by providing ancillary services.  

To outline our framework, we start with the following formulation of the OPF problem (we present some extensions later in Section~\ref{sec:extensionVPP}):
\begin{equation}\label{eq:OPFProb}
    \begin{aligned}
        \min_{\vecV \in \mathbb{R}^M, \vecu \in \mathbb{R}^{2G}} \quad & C_v(\vecV) + C_p(\vecu) \\
        \textrm{s.t.} \quad
         & \underline{V}\leq \nu_i \leq \bar{V} \qquad \quad \  \forall i \in \mathcal{M}\\
         & \nu_i = H_i(\vecu;\vecpl,\vecql)\quad \forall i \in \mathcal{M}\\
          & (p_i,q_i) \in \mathcal{U}_i \qquad \quad \,\,\, \forall i \in \mathcal{G}
    \end{aligned}
\end{equation}
where the functions $C_v: \mathbb{R}^{M} \rightarrow \mathbb{R}$ and  $C_p : \mathbb{R}^{2G} \rightarrow \mathbb{R}$  have locally Lipschitz continuous gradients, $\underline{V}$ and $\bar{V}$ are predefined voltage bounds that the operator wants to enforce at nodes $i \in \mathcal{M}$, $H_i(\vecu;\vecpl,\vecql)$ is the $i$th component of the function $H(\vecu;\vecpl,\vecql)$ (specifying the voltage magnitude $\nu_i$). 
We note that~\eqref{eq:OPFProb} can be equivalently re-written as:
\begin{equation}\label{eq:OPFProb2}
    \begin{aligned}
        \min_{ \vecu \in \mathbb{R}^{2G}} \quad & C_v(H(\vecu;\vecpl,\vecql) ) + C_p(\vecu) \\
        \textrm{s.t.} \quad
         & \underline{V}\leq H_i(\vecu;\vecpl,\vecql) \leq \bar{V} \quad \forall i \in \mathcal{M} \\
          &  (p_i,q_i) \in \mathcal{U}_i \qquad \qquad \quad \,\,\, \forall i \in \mathcal{G}
    \end{aligned}
\end{equation}
where $\vecu$ is the only optimization variable.  
Hereafter, we assume the set $\mathcal{U}_i$ can be expressed as
\begin{align}
    \mathcal{U}_i = \{(p_i,q_i) \in \mathbb{R}^2: \ell_i( p_i,q_i) \leq \zero_{n_{c_i}}\}
\end{align}
where $\ell_i: \mathbb{R}^2 \rightarrow \mathbb{R}^{n_{c_i}}$ is a vector-valued function modeling power limits, and the inequality is taken entry-wise.  For example,
if the $i$th DER is an inverter-interfaced controllable renewable source, then $\ell_i(p_i,q_i) = [p_i^2+q_i^2 - s_{n,i}^2, p_i - p_{\text{max},i}, -p_i]^\top$, where $s_{n,i}$ and $p_{\text{max},i}$ denote the inverter rated size and the maximum available active power, respectively; that is, $\mathcal{U}_i = \{(p_i,q_i) \in \mathbb{R}: p_i^2+q_i^2 \leq s_{n,i}^2, p_i\leq p_{\text{max},i}, p_i\geq 0 \}$. Moreover, we denote as 
\begin{align}
\mathcal{F} := \{\vecu: \underline{V}\leq H_i(\vecu;\vecpl,\vecql) \leq \bar{V}, \, \forall i \in \mathcal{M}, \vecu \in \mathcal{U}\}
\end{align}
the feasible set of~\eqref{eq:OPFProb2}. We impose the following assumption on~\eqref{eq:OPFProb2}, which is typical in the AC OPF context. 


\begin{assumption}[Regularity of isolated solutions]
\label{as:openloop}
    Assume that~\eqref{eq:OPFProb2} is feasible and  let $\vecu^*$ be a local minimizer and an isolated Karush–Kuhn–Tucker (KKT)
    point for~\eqref{eq:OPFProb2}, for given $\vecpl,\vecql$. Assume that the following hold: 

    \noindent \textit{i)}
    Strict complementarity slackness~\cite{fiacco1976sensitivity} and the linear independence constraint qualification (LICQ)~\cite{hauswirth2018generic} hold at $ \vecu^*$.
    
    \noindent \textit{ii)}
    The maps $\vecu \mapsto C_p(\vecu)$, $\vecu \mapsto C_v(H(\vecu;\vecpl,\vecql) ) $ and $\vecu\mapsto H(\vecu;\vecpl,\vecql)$ are twice continuously differentiable over some open neighborhood $\mathcal{B}(\vecu^*,r_1):=\{\vecu : \|\vecu-\vecu^*\|<r_1 \}$ of $\vecu^*$, and their Hessian matrices are positive semi-definite at $\vecu^*$.

    \noindent \textit{iii)}
    The Hessian $\nabla^2  C_p(\vecu^*)$ is positive definite.
     \hfill $\Box$
\end{assumption}

Assumption~\ref{as:openloop} imposes some mild regularity assumptions on a neighborhood of a strict locally optimal solution. If~\eqref{eq:OPFProb2} is formulated based on the linearized AC power flow equations~\cite{bolognani2014distributed,dall2016optimal} and the cost is strongly convex, then  Assumption~\ref{as:openloop} is satisfied. For the most general nonlinear AC OPF problem~\cite{Tang17}, Assumption~\ref{as:openloop} is supported by the results of~\cite{hauswirth2018generic}, where LICQ is investigated. 

Problem~\eqref{eq:OPFProb2} can be solved using traditional optimization methods for nonlinear programs. However, \emph{(c1)} these batch methods require collecting measurements of all the \emph{non-controllable} powers  $\vecpl, \vecql$ in a distribution network that is known to be historically measurement-scarce~\cite{cheng2023survey};  moreover, \emph{(c2)} the time required to collect the measurements of the non-controllable powers (if available in real time) and run an iterative method to convergence may be long compared to the fast changing conditions of a modern distribution system~\cite{taylor2016power,kroposki2020autonomous}.
Although recent work on neural networks for AC OPF can alleviate the computational burden (see, e.g., the representative works~\cite{baker2020emulating,nellikkath2022physics}), they still require measurements of all the non-controllable powers $\vecpl, \vecql$ as in \emph{(c1)}, and they often rely on \emph{heuristics} to return a feasible solution. Motivated by the challenges \emph{(c1)}--\emph{(c2)} and by the need to generate power setpoints that ensure satisfaction of voltage limits even under uncertain and time-varying operational setups, in this paper we seek to solve to the following problem. 

\begin{problem}\label{prob:Problem1}
    Design an online feedback-optimization algorithm that drives the DERs' power setpoints $\vecu$ to solutions of the problem \eqref{eq:OPFProb2}, while ensuring that voltage constraints are always met. The feedback optimization method should use measurements of the voltages instead of requiring knowledge of the non-controllable powers $\vecpl,\vecql$. \hfill $\Box$
\end{problem}

We note that in Problem~\ref{prob:Problem1} we focus on voltage measurements because~\eqref{eq:OPFProb2} includes voltage constraints; if~\eqref{eq:OPFProb2} is modified to include cost and constraints associated with power flows or currents, then the feedback optimization method would require measurements of those quantities too~\cite{bernstein2019real}.

From a practical standpoint, in this paper we focus on   feedback-optimization algorithms that are implemented in a centralized unit; for instance, these algorithms can be integrated into a DERMS for distribution operators. 
We also assume that the unit implementing our feedback-based online algorithm has access to synchronized  voltage measurements at nodes where voltage constraints are enforced, and can transmit  new power setpoints to the DERs. This can be done by leveraging  existing communication and metering infrastructure or through a SCADA system.

\section{Safe OPF pursuit}\label{sec:safeopfpursuit}

\subsection{Feedback-based online algorithm}

To solve our regulation problem, 
we propose the following  feedback-based algorithm: 
\begin{align}
    & \dot{\vecu} = \eta F_{\beta}(\vecu, \tilde \vecV) \label{eq:controller} \\
    & F_{\beta}(\vecu, \tilde \vecV) := \nonumber\\
    & \arg \min_{\boldsymbol{\theta} \in \mathbb{R}^{2G}}  \|\boldsymbol{\theta} + \nabla C_p(\vecu) + J_H(\vecu;\vecpl,\vecql)^\top\nabla C_v(\tilde \vecV)\|^2 \nonumber  \\
    & \hspace{1cm}  \textrm{s.t.}  -\nabla H_i(\vecu;\vecpl,\vecql)^\top\boldsymbol{\theta} \leq -\beta \left(\underline{V}-\tilde \nu_i\right) \ \forall i \in \mathcal{M} \label{eq:controller_problem} \\
    & \hspace{1.5cm} \nabla H_i(\vecu;\vecpl,\vecql)^\top\boldsymbol{\theta} \leq -\beta \left(\tilde \nu_i -\bar{V}\right) \quad \forall i \in \mathcal{M} \nonumber \\        
    &  \hspace{1.5cm} J_{\ell_i}(p_i,q_i)^\top \boldsymbol{\theta} \leq - \beta \ell_i(p_i,q_i) \qquad  \qquad \forall i \in \mathcal{G} \nonumber 
\end{align}
where $\tilde \nu_i$ is a measurement of $|v_i|$ at node $i$, $J_H(\vecu;\vecpl,\vecql)$ is the Jacobian matrix of $H(\vecu;\vecpl,\vecql)$, $ \nabla H_i(\vecu;\vecpl,\vecql) = [\{(J_H(\vecu;\vecpl,\vecql))_{i,j}\}_{j\in\mathcal{G}}]^\top$ is a $2G\times 1$ vector collecting the entries of $J_H(\vecu;\vecpl,\vecql)$ in the $i$th row and columns corresponding to nodes in $\mathcal{G}$, $J_{\ell_i}(p_i,q_i)$ is the Jacobian of $(p_i,q_i) \mapsto \ell_i(p_i,q_i)$,  $\beta > 0$ is a design parameter, and $\eta > 0$ is the controller gain and is a design parameter. For given $\vecu$ and $\tilde{\vecV}$, problem~\eqref{eq:controller_problem} is a convex quadratic program (QP) with a strongly convex cost; it can be efficiently solved using standard  or high-performance embedded  solvers for QPs, e.g.,~\cite{banjac2017embedded}. 

The online feedback optimization algorithm~\eqref{eq:controller} is inspired by CBFs methods~\cite{ames2019control} and the safe gradient flow in~\cite{allibhoy2023control}; we provide more details on the CBF-based design in Appendix~\ref{sec:design}. In particular,~\eqref{eq:controller} is an approximation of the projected gradient flow 
$\dot \vecu = \text{proj}_{T_\mathcal{F}(\vecu)}\{-\nabla C_p(\vecu) - J_H^\top\nabla C_v(H(\vecu;\vecpl, \vecql))\}
$, where $T_\mathcal{F}(\vecu)$ is the tangent cone of $\mathcal{F}(\vecu)$ at $\vecu$; in fact, one can show~\cite[Prop. 4.4]{allibhoy2023control} that $\lim_{\beta \rightarrow \infty} F_\beta(\vecu,H(\vecu;\vecpl, \vecql)) = \text{proj}_{T_\mathcal{F}(\vecu)}\{-\nabla C_p(\vecu) - J_H^\top\nabla C_v(H(\vecu;\vecpl, \vecql))\}$. 

\begin{figure}
    \centering
    \includegraphics[scale=0.9]{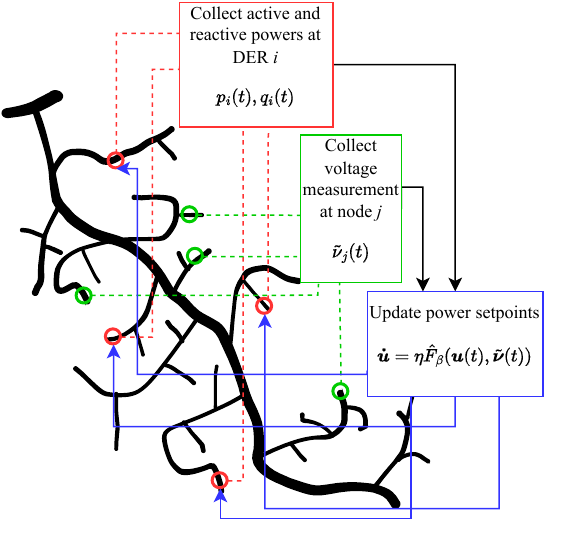}
    \caption{Closed-loop implementation of the proposed online feedback optimization algorithm. A central unit (blue box) receives measurements of voltages (from measurement units, green box) and of the DERs' output powers from inverters (red box); based on these measurements, it updates the DERs' setpoints based on the proposed controller $\dot{\vecu}(t) = \eta \hat{F}_{\beta}(\vecu(t), \tilde{\vecV}(t))$. Once the setpoints $\boldsymbol{u}(t)$ are computed, the central unit transmits $\boldsymbol{u}(t)$ to the DERs' inverters. Through this closed-loop scheme, the proposed controllers drive the distribution system to solutions of the AC OPF problem~\eqref{eq:OPFProb2}.}
    \label{fig:proposed_arch}
    \vspace{-.4cm}
\end{figure}

The algorithm~\eqref{eq:controller} is designed to  steer the power setpoints of the DERs $\vecu$ to optimal solutions of the AC OPF, while continuously guaranteeing feasibility (i.e., satisfaction of voltage limits). As shown in Figure~\ref{fig:proposed_arch},~\eqref{eq:controller}  effectively acts as a feedback controller by replacing the voltage \emph{model} $H(\vecu;\vecpl, \vecql)$ with \emph{measurements} $\tilde \vecV$ of  the voltage magnitudes that automatically satisfy the power flow equations~\cite{Tang17, molzahn2017survey}. This is a key modification that allows one to avoid collecting measurements of $\vecpl, \vecql$~\cite{bolognani2014distributed,  dall2016optimal,bernstein2019real}. However, we note that~\eqref{eq:controller} requires the computation of the Jacobian matrix of $H(\vecu;\vecpl,\vecql)$. One cannot derive an explicit formulation of the Jacobian matrix of $H(\vecu;\vecpl,\vecql)$ as this map does not have an analytical formulation. Therefore, we modify~\eqref{eq:controller} as follows: 
\begin{align}
     & \hspace{-.2cm} \dot{\vecu} = \eta \hat{F}_{\beta}(\vecu, \tilde \vecV) \label{eq:controller_approx} \\
     & \hspace{-.2cm}  \hat{F}_{\beta}(\vecu, \tilde \vecV) := \arg \min_{\boldsymbol{\theta} \in \mathbb{R}^{2G}}  \|\boldsymbol{\theta} + \nabla C_p(\vecu) + J_{\hat{H}}^\top\nabla C_v(\tilde \vecV)\|^2 \nonumber  \\
       & \hspace{1.7cm}  \textrm{s.t.}  -\nabla \hat{H}_i^\top\boldsymbol{\theta} \leq -\beta \left(\underline{V}-\tilde \nu_i\right) \,\,\, \forall i \in \mathcal{M} \label{eq:controller_problem_approx} \\
        & \hspace{2.55cm} \nabla \hat{H}_i^\top\boldsymbol{\theta} \leq -\beta \left(\tilde \nu_i -\bar{V}\right) \,\, \forall i \in \mathcal{M} \nonumber \\        
        &  \hspace{2.55cm} J_{\ell_i}(\vecu)^\top \theta \leq - \beta \ell_i(p_i,q_i) \quad  \,\,\,\, \forall i \in \mathcal{G} \nonumber 
\end{align}
where $J_{\hat{H}}$ and $\{\hat{H}_i\}_{i \in \mathcal{M}}$ are estimates or approximations of $J_{H}$ and $\{H_i\}_{i \in \mathcal{M}}$, respectively. These estimates can be obtained using online estimation methods (see, e.g.,~\cite{picallo2022adaptive,ospina2023data}), or they can be computed based on a linear approximation of the AC power flow equations of the form
\begin{equation}
\label{eq:linearmodel}
        \hat{H}_n(\vecu;\vecpl,\vecql) = \sum_{i\in\mathcal{G}}(r_{n,m(i)}p_i + b_{n,m(i)} q_i) + c_n(\vecpl,\vecql)         
\end{equation}
$n \in \mathcal{N}$, where the coefficients $\{r_{n,m(i)}, b_{n,m(i)}\}_{i\in \mathcal{G}}$ can be found as explained in e.g.,~\cite{bolognani2014distributed,bolognani2015existence,bernstein2018load, LinearApprox2, kekatos2015voltage} (note that linear approximations of the power flow equations are available for both single-phase and unbalanced multi-phase distribution networks). We will provide more remarks on the linear approximation shortly in  Section~\ref{sec:linear}.

The proposed feedback-based SGF is summarized in Algorithm~\ref{alg:safe} and illustrated in Fig.~\ref{fig:proposed_arch}.

\begin{algorithm}[h!]
\caption{\emph{Feedback-based safe gradient flow}}
\label{alg:safe}
\textbf{Initialization}: Compute $J_{\hat{H}}$ and $\{\hat{H}_i, i \in \mathcal{M}\}$. Set $\beta > 0$, $\eta > 0$. 

\textbf{Real-time operation}: for $t \geq 0$, repeat:

\quad [S1a] Measure output powers $\{ p_i(t),q_i(t), i \in \mathcal{G}\}$ 

\quad [S1b] Measure voltages $\{\tilde{\nu}_i(t), i \in \mathcal{M}\}$

\quad [S2a] Update power setpoints via $\dot{\vecu}(t) = \eta \hat{F}_{\beta}(\vecu(t), \tilde{\vecV}(t))$

\quad [S2b] Implement setpoints $\vecu(t)$  

\quad Go to [S1a] and [S1b]

\end{algorithm}

In terms of implementation of the Algorithm~\ref{alg:safe}, we highlight the following practical aspects: 
\begin{itemize}
    \item The main step [S2a] is performed at a central unit (i.e., the blue box in Fig.~\ref{fig:proposed_arch}). This central unit can be integrated, for example, into a DERMS or an advanced distribution management system for distribution operators. After performing step [S2a], the central unit sends the updated setpoints to the DERs' inverters.  
    \item Step [S1a] is performed at the DERs (a DER is represented by a red box in Fig.~\ref{fig:proposed_arch}); the inverters measure the output powers $\{ p_i(t),q_i(t), i \in \mathcal{G}\}$  and send the measurements to the central unit.  The inverters also implement step [S2b] after they receive the setpoints from the central unit. 
    \item The safe gradient flow~\eqref{eq:controller_approx} relies on measurements of the voltages at the network locations $\mathcal{M}$, as required in the step [S1b]. It is assumed that those measurements are obtained in real time using sensing devices communicating with the centralized controller, e.g., $\mu$PMUs~\cite{8289180} or through the advanced metering infrastructure (a meter is represented by a green box in Fig.~\ref{fig:proposed_arch}).
\end{itemize}
%
%
In practice, the proposed measurement-based SGF~\eqref{eq:controller_approx} can  be implemented with discretization (similar to well-established CBF-based methods~\cite{ames2019control}). The discretization interval depends on the time required to collect voltage measurements and to solve the QP. Linearly-constrained convex QP programs are known to be solved efficiently (e.g, in milliseconds) by both existing open-source solvers (such as IPOPT) and commercial solvers. Synchronized voltage measurements can be obtained via SCADA at fast scale (i.e., second or sub-second level)~\cite{cheng2023survey,8289180}.


\begin{remark}[\textit{Pseudo-measurements}]
{\rm
Our framework is applicable to the case where the system operator may utilize a mix of actual voltage measurements and pseudo-measurements~\cite{angioni2015impact}. 
For example, suppose that the system operator can measure voltages at some nodes $\mathcal{M}_{\text{meter}}\subset \mathcal{M}$ and relies on pseudo-measurements at the other nodes. Then,  
\begin{equation}
\Tilde{\nu}_i=\begin{cases}
       H_i(\vecu;\vecpl,\vecql)+n_i,~~~i\in \mathcal{M}_{\text{meter}} \\
       H_{i,\text{pseudo}}(\vecu;\vecpl,\vecql),~~~i\in  \mathcal{M}\setminus \mathcal{M}_{\text{meter}} 
    \end{cases}
    \label{eq:pseudo-measurement}
    \end{equation}
where $n$ is a bounded measurement noise, and $H_{i,\text{pseudo}}(\vecu;\vecpl,\vecql)$ represents a model used to generate the pseudo-measurements (i.e., using a state-estimator). 
\hfill $\Box$
}
\end{remark}

\begin{remark}[\textit{Measurement of setpoints}]
{\rm
If the embedded controllers of inverters are guaranteed to implement the power setpoints, in principle, the step [S1a] in Algorithm~\ref{alg:safe} is not needed. However, the operator may want to measure current setpoints $\{p_i(t),q_i(t), i \in \mathcal{G}\}$ for verification purposes and to monitor the state of the DERs' inverters. 
\hfill $\Box$
}
\end{remark}


\subsection{Remarks on the linear model}
\label{sec:linear}

In the following, we comment on the linear approximation used in this paper. We start with the power flow equations~\eqref{eq:pfequations} that we linearize around a given voltage profile $\bar{\vecv} = [\bar{v}_1,...,\bar{v}_N]^\top$. Let us consider $\boldsymbol{d} \in \mathbb{R}^{N}$ capturing the deviations around the linearization point. We have:
\begin{multline}\label{eq:pf_smalldev}
    \vecs = \diag(\bar{\vecv}+\boldsymbol{d})\left(\smally^*v_0^* + \capitalY^*\bar{\vecv}^*\right) + \diag(\boldsymbol{d})\left(\capitalY^*\bar{\vecv}^*\right) +\\ \diag(\boldsymbol{d})\left(\capitalY^*\boldsymbol{d}^*\right)  .
\end{multline}
If we discard the second-order terms $\diag(\boldsymbol{d})\left(\capitalY^*\boldsymbol{d}^*\right)$, and considering the following choice for the nominal voltage profile:
\begin{equation}
    \bar{\vecv} = -\capitalY^{-1}\smally v_0, 
\end{equation}
equation~\eqref{eq:pf_smalldev} becomes:
\begin{equation}\label{eq:lin_pfeq}
    \diag(\bar{\vecv}^*)\capitalY \boldsymbol{d} = \vecs^*.
\end{equation}
Let $\bar{\boldsymbol{\rho}} \in \mathbb{R}^N$ be the vector collecting the magnitudes of voltages $\bar{\vecv}$, and define  $\bar{\boldsymbol{a}} := \{\cos(\bar{\theta}_n)\}_{n\in\N}$, $\bar{\boldsymbol{b}} := \{\sin(\bar{\theta}_n)\}_{n\in\N}$ with $\bar{\theta}_i$ the angle of the nominal voltage $\bar{v}_i$.
A solution of~\eqref{eq:lin_pfeq} can be expressed as $\boldsymbol{d} = \capitalY^{-1}\diag^{-1}(\bar{\vecv}^*)\vecs^*$. Expanding this expression, and defining matrices:
\begin{equation}
    \begin{aligned}
        \bar{\boldsymbol{R}} = \boldsymbol{Z}_R\diag(\bar{\boldsymbol{a}})(\diag(\bar{\boldsymbol{\rho}}))^{-1} -\boldsymbol{Z}_I\diag(\bar{\boldsymbol{b}})(\diag(\bar{\boldsymbol{\rho}}))^{-1}\\
        \bar{\boldsymbol{B}} = \boldsymbol{Z}_I\diag(\bar{\boldsymbol{a}})(\diag(\bar{\boldsymbol{\rho}}))^{-1} +\boldsymbol{Z}_R\diag(\bar{\boldsymbol{b}})(\diag(\bar{\boldsymbol{\rho}}))^{-1},       
    \end{aligned}
\end{equation}
where $\boldsymbol{Z}_R := \Re\{\capitalY^{-1}\}$ and $\boldsymbol{Z}_I := \Im\{\capitalY^{-1}\}$, one can write:
\begin{equation}
    \vecv \approx \left(\bar{\boldsymbol{R}} + j\bar{\boldsymbol{B}}\right)\vecp_{\mathrm{net}} + \left(\bar{\boldsymbol{B}} - j\bar{\boldsymbol{R}}\right)\vecq_{\mathrm{net}} + \bar{\vecv}.
\end{equation}
If the entries of $\bar{\vecv}$ dominate those in $\boldsymbol{d}$, then $\boldsymbol{\rho}+\Re\{\boldsymbol{d}\}$ serves as a first order approximation for the voltage magnitudes. Thus, one can write:
\begin{equation}\label{eq:H_developed}
    \hat{H}(\vecu;\vecpl,\vecql) := \bar{\boldsymbol{R}}(\boldsymbol{\Gamma}_R\vecu + \vecpl)+\bar{\boldsymbol{B}}(\boldsymbol{\Gamma}_B\vecu + \vecql)+\bar{\boldsymbol{\rho}},
\end{equation}
with $\boldsymbol{\Gamma}_R \in \mathbb{R}^{N\times 2G}$ and $\boldsymbol{\Gamma}_B \in \mathbb{R}^{N\times 2G}$ matrices filled with 0 and 1 such that $\boldsymbol{\Gamma}_R\vecu + \vecpl = \vecp_{\mathrm{net}}$ and $\boldsymbol{\Gamma}_B\vecu + \vecql = \vecq_{\mathrm{net}}$. Notice that equation~\eqref{eq:H_developed} can be written as in~\eqref{eq:linearmodel}. Effectively, the approximate Jacobian $J_{\hat{H}} =\bar{\boldsymbol{R}}\boldsymbol{\Gamma}_R+\bar{\boldsymbol{B}}\boldsymbol{\Gamma}_B$ 
no longer depends on $\vecu$ and the non-controllable powers $\vecpl,\vecql$; accordingly, it does not need to be re-computed when running~\eqref{eq:controller_approx}.

\begin{remark}[\textit{Validity of the linear approximation}]{\rm
    The linear model is based on the bus admittance matrix $\capitalY$ and has constant matrices. This approximation is accurate for lightly loaded systems~\cite{bolognani2015fast}. For heavily loaded system,~\cite{ortmann2020experimental,dall2016optimal} showed that feedback-based methods are robust against model mismatch because of the closed-loop implementation; this feature is also  pointed out in~\cite{molzahn2017survey}, and our analysis in the next section will characterize this robustness. The bus admittance matrix may be hard to obtain because it requires knowledge of the feeder characteristics, \textit{i.e.}, the line impedances, and the network configuration. However, one can assume that the system operator can obtain some estimates. Furthermore, in the case of network reconfiguration, the bus admittance matrix changes, leading to an incorrect linear model. However, this is not a frequent event and the system operator can update its linear model approximation when a reconfiguration occurs. 
    \hfill $\Box$}
\end{remark}

In the next section, we analyze the convergence and stability properties of the proposed feedback-based SGF~\eqref{eq:controller_approx}

\subsection{Stability analysis and constraint satisfaction guarantees}
\label{sec:stability}

In our technical analysis, we make use of the following assumptions. The assumptions are stated for given values of the non-controllable powers $\vecpl,\vecql$. 

\begin{assumption}[\textit{Jacobian errors}] 
\label{as:bound_approx_error}
    $\exists ~ E_h < + \infty, E_J < + \infty$ such that $\|\hat{H}(\vecu;\vecpl,\vecql)-{H}(\vecu;\vecpl,\vecql)  \| \leq E_h$ and $\|J_{\hat{H}}(\vecu) -J_H(\vecu)  \| \leq E_J $ for any  $\vecu \in \mathcal{B}(\vecu^*,r_1)$. \hfill $\Box$
\end{assumption}

\begin{assumption}[\textit{Measurement errors}] 
\label{as:bound_meas_error}
$\exists ~ E_M < + \infty$ such that $\|\tilde \vecV - \vecV\| \leq E_M$. \hfill $\Box$
\end{assumption}

Assumptions~\ref{as:bound_approx_error}-\ref{as:bound_meas_error} are motivated by the following observations: (i) the linear map error $\|\hat{H}(\vecu;\vecpl,\vecql)-H(\vecu;\vecpl,\vecql)\|$ is bounded and small in a neighborhood of the optimizer (as confirmed in Fig.~\ref{fig:Maperror} in our numerical results, and by the analytical findings in~\cite{bolognani2015existence, LinearApprox2}), and (ii) in realistic monitoring and SCADA systems, the measurement of the voltage magnitudes are affected by a small (or even negligible) error.

In our analysis, we view  \eqref{eq:controller_approx}  as a perturbed version of \eqref{eq:controller}. To begin with, 
we have the following result. 

\begin{lemma}[\textit{KKT and equilibrium}] 
\label{lem:equilibrium}
Consider the problem~\eqref{eq:OPFProb2} satisfying  Assumptions \ref{as:steadyStateMap}-\ref{as:openloop}. There exists $\boldsymbol{\mu}^*$ such that $(\vecu^*,\boldsymbol{\mu}^
*)$ is a KKT point for \eqref{eq:OPFProb2} if and only if $\vecu^*$ is an equilibrium of  $\dot{\vecu} = \eta F_{\beta}(\vecu, H(\vecu;\vecpl,\vecql))$.  \hfill $\Box$  
\end{lemma}

Before analyzing the stability of the proposed feedback-based SGF, we provide some notation and intermediate results that will be used in the proof of our main result.

Let $\Omega := J_{\hat{H}}(\vecu) - J_H(\vecu)$  and denote by $\omega_i$ the $i$-th row of $\Omega$. Moreover, let $\mathbf{e} := \tilde \vecV - \vecV$ denote the measurement errors. Then, define $\bar{F}_\beta(\vecu,\Omega,\mathbf{e})$ as
\begin{align*}   
&  \bar{F}_\beta(\vecu,\Omega,\mathbf{e}) \\
&:= \arg \min_{\boldsymbol{\theta}} \|\boldsymbol{\theta} \!+\! \nabla C_p(\vecu) \!+\! (J_H(\vecu)\!+\!\Omega)^\top\nabla C_v(\vecV+\mathbf{e})\|^2\\
      & \hspace{1.cm}  \textrm{s.t.}  -(\nabla {H}_i(\vecu) +\omega_i)^\top\boldsymbol{\theta} \leq -\beta \left(\underline{V}- \nu_i - e_i\right) \,\, \forall i \in \mathcal{M} \nonumber \\
        & \hspace{1.85cm} (\nabla {H}_i(\vecu) +\omega_i)^\top\boldsymbol{\theta} \leq -\beta \left(\nu_i + e_i -\bar{V}\right) \,\, \forall i \in \mathcal{M} \nonumber \\        
        &  \hspace{1.85cm} J_{\ell_i}(\vecu)^\top \theta \leq - \beta \ell_i(p_i,q_i) \qquad \qquad \quad \quad \,\,\,\,\,\, \forall i \in \mathcal{G} \nonumber        
\end{align*}
where $\vecV = H(\vecu;\vecpl,\vecql)$. Note that $F_\beta(\vecu,H(\vecu;\vecpl,\vecql)) = \bar{F}_\beta(\vecu,\zero,\zero)$ and $\hat{F}_{\beta}(\vecu, \tilde \vecV) = \bar{F}_\beta(\vecu,J_{\hat{H}}(\vecu) - J_H(\vecu),\tilde \vecV - \vecV)$. Let $\mathcal{E}_J := \{\Omega: \|\Omega\| \leq E_J\}$ and $\mathcal{E}_M := \{\mathbf{e}: \|\mathbf{e}\| \leq E_M\}$ for brevity. We make the following assumption on~$\bar{F}_\beta$. 

\begin{assumption}[\textit{Regularity}] 
\label{as:Regularity}
For any $\vecu  \in \mathcal{B}(\vecu^*,r_1)$, and any $\Omega$ and $\mathbf{e}$ satisfying Assumptions~\ref{as:bound_approx_error}-\ref{as:bound_meas_error}, the problem~\eqref{eq:controller_problem_approx} is feasible,  and satisfies the Mangasarian-Fromovitz Constraint Qualification  and the constant-rank condition~\cite{liu1995sensitivity}. 
 \hfill $\Box$
\end{assumption}

Since the constraints in the problem defining $\bar{F}_\beta(\vecu,\Omega,\mathbf{e})$ (and, hence, our safe gradient flow~\eqref{eq:controller_approx}) are based on techniques from CBFs~\cite{ames2019control,allibhoy2023control}, Assumption~\ref{as:Regularity} guarantees that there always exists a direction for the setpoints to satisfy the constraints of the OPF. Moreover, this assumption allows us to derive the following result.  

\begin{lemma}[\textit{Lipschitz continuity}] 
\label{lem:lipschitz}
Let Assumption~\ref{as:Regularity} hold, and assume that $\vecu \mapsto C_p(\vecu)$, $\vecV \mapsto C_v(\vecV) $ are twice continuously differentiable over $\mathcal{B}(\vecu^*,r_1)$ and $\mathcal{V}:= \{ \vecV \in \mathbb{R}^M:  \underline{V}\leq \nu_i + e_i \leq \bar{V}, \, \forall i \in \mathcal{M}, \vecV = H(\vecu;\vecpl,\vecql), \|\mathbf{e}\| \leq E_M,  \vecu \in \mathcal{B}(\vecu^*,r_1) \}$, respectively. Then: 

\noindent (i) For any $\Omega \in \mathcal{E}_J$ and $\mathbf{e} \in \mathcal{E}_M $, $\vecu \mapsto \bar{F}_\beta(\vecu,\Omega,\mathbf{e})$ is locally Lipschitz at $\vecu$, $\vecu \in \mathcal{B}(\vecu^*,r_1)$. 

\noindent (ii) For any $\vecu \in \mathcal{B}(\vecu^*,r_1)$ and $\Omega \in \mathcal{E}_J$, $\mathbf{e} \mapsto \bar{F}_\beta(\vecu,\Omega,\mathbf{e})$ is  Lipschitz with constant $\ell_{F_{v}} \geq 0$  over $\mathcal{E}_M$.

\noindent (iii) For any $\vecu \in \mathcal{B}(\vecu^*,r_1)$ and $\mathbf{e} \in \mathcal{E}_M $, $\Omega \mapsto \bar{F}_\beta(\vecu,\vecV,\Omega,\mathbf{e})$ is  Lipschitz with constant $\ell_{F_{J}} \geq 0$  over $\mathcal{E}_J$. \hfill $\Box$
\end{lemma}

Lemma~\ref{lem:lipschitz} follows from~\cite[Theorem 3.6]{liu1995sensitivity}, and by the compactness of the sets $\mathcal{E}_M$ and $\mathcal{E}_J$.  
This result ensures existence and uniqueness of solutions for the proposed feedback-based safe gradient flow~\cite[Ch.~3]{khalil2002nonlinear}.

Our main stability result critically relies on these results. Before stating it,  we introduce some useful quantities that play a role in the main result; in particular, they are related to local properties of $F_\beta(\vecu,H(\vecu;\vecpl,\vecql)) $. Recall that $\vecu^*$ is the local optimizer of \eqref{eq:OPFProb2}.
We define $\vecV^*:=H(\vecu^*;\vecpl,\vecql)$, 
$E:=\frac{\partial F_\beta (\vecu,H(\vecu; \vecpl, \vecql))}{\partial \vecu}\mid_{\vecu=\vecu^*}$, 
$e_1 :=-\lambda_{\max}(E)$, 
and 
$e_2 :=-\lambda_{\min}(E)$. 
Then, we can write the dynamics as $F_\beta(\vecu,H(\vecu; \vecpl, \vecql)) = E (\vecu-\vecu^*) + \hat{g}(\vecu)$, where $\hat{g}(u)$ satisfies $\|\hat{g}(\vecu)\|\leq L \|\vecu-\vecu^*\|^2$, $\forall \vecu\in \mathcal{B}(\vecu^*,r_2)$, for some $L> 0$ and $r_2>0$ (see~\cite{khalil2002nonlinear}). Define $r:=\min\{r_1,r_2\}$ and
\begin{align*}
s_{\min}:=
\begin{cases}
0, & \text{ if } r\geq \frac{e_1}{L}, 
\\ 
1-\frac{r L}{e_1}, & \text{ if } r<\frac{e_1}{L} .    
\end{cases}
\end{align*}
Since $\mathcal{U}$ is compact, $J_H(\vecu)$ is Lipschitz on $\mathcal{U}$ with constant $\ell_{H}$. 
We are now ready to state the main stability result for~\eqref{eq:controller_approx}.

\begin{theorem}[\textit{Practical local exponential stability}] 
\label{thm:stability}
 Consider the OPF problem~\eqref{eq:OPFProb2} satisfying  Assumptions \ref{as:steadyStateMap}-\ref{as:openloop}, a linear map $\hat{H}$ satisfying  Assumption \ref{as:bound_approx_error},  measurements $\tilde{\vecV}$ satisfying  Assumption \ref{as:bound_meas_error}, and the controller~\eqref{eq:controller_approx} satisfying Assumption~\ref{as:Regularity}. Let $\vecu(t)$, $t \geq t_0$,  be the unique trajectory of~\eqref{eq:controller_approx}.  Assume that the set $ \mathcal{S}:=\left\{s:s_{\min}<s\leq 1,~e_1^{-3} e_2 L(\ell_{F_J}E_J+\ell_{F_v} E_M   )<s-s^2\right\}$ is not empty. Then, for any $s\in\mathcal{S}$, it holds that 
\begin{align}
 &\|\vecu(t)-\vecu^*\|
 \leq \sqrt{\frac{e_2}{e_1}} e^{- e_1\eta s(t-t_0)}\|\vecu(t_0)-\vecu^*\| \nonumber \\
 & \hspace{.9cm} + \frac{e_2(\ell_{F_J}E_J+\ell_{F_v} E_M   )}{ s e_1^2} \left(1-e^{- e_1\eta s(t-t_0)}     \right) \label{eq:transient_u},
\end{align}
for any initial condition $\vecu(t_0)$ such that $\|\vecu(t_0)-\vecu^*\|\leq \sqrt{\frac{e_1}{e_2}}\frac{e_1}{L}(1-s)$. 
\hfill $\triangle$
\end{theorem}
The proof of the result is provided in the Appendix~\ref{ap:proof}. The assumption that  $\mathcal{S}$ is not empty is necessary to guarantee that the trajectory of $\vecu(t)$  never exits the region of attraction of the optimizer $\vecu^*$. We can notice that the first term on the right-hand-side of~\eqref{eq:transient_u} decays over time; the second term models the effect of the measurement errors and of the errors in the computation of the Jacobian. In particular, we can notice that, as $t\to+\infty$, the right-hand side of~\eqref{eq:transient_u} becomes
\begin{align}
\label{eq:asymp} 
\lim_{t \rightarrow + \infty} \|\vecu(t)-\vecu^*\| \leq s^{-1} e_1^{-2} e_2(\ell_{F_J}E_J+\ell_{F_v} E_M   ) .
\end{align}
The asymptotic error can be reduced by increasing the accuracy in the measurement of the voltages (i.e., reducing $E_M$), or allocating more computational power to compute the Jacobian of the power flow equations (i.e., reducing $E_J$). The following result characterizes the feasibility of the solution $u(t)$.

\begin{lemma}[\textit{Practical forward invariance}]
\label{lem:pratical_forward}
Let  the conditions in Theorem~\ref{thm:stability} be satisfied, and let $\vecu(t)$, $t \geq t_0$,  be the unique trajectory of~\eqref{eq:controller_approx}, and $\vecV(t)$ be the corresponding voltage magnitudes. Define the set 
\begin{align}
\hspace{-.2cm} \mathcal{F}_e := \{\vecu: \vecu \in \mathcal{U},  \underline{V}_e \leq H_i(\vecu;\vecpl,\vecql)  \leq \bar{V}_e, \forall i \in \mathcal{M}\}
\end{align}
with $\underline{V}_e := \underline{V}-E_M-2E_{\hat{H}}$, $\bar{V}_e := \bar{V}+E_M+2E_{\hat{H}}$, and 
$E_{\hat{H}}:=\max_{\vecu\in \mathcal{U}} \| \hat{H}(\vecu;\vecpl,\vecql)-H(\vecu;\vecpl,\vecql)\|$. Then, the feedback-based SGF~\eqref{eq:controller_approx} renders a set $ \mathcal{F}_s$, with $\mathcal{F} \subseteq  \mathcal{F}_s \subseteq \mathcal{F}_e$, forward invariant.
  \hfill  $\Box$
\end{lemma}

The proof is provided in Appendix~\ref{ap:pratical_forward}. Lemma~\ref{lem:pratical_forward} establishes forward invariance of a set $\mathcal{F}_s$, which is a subset of $\mathcal{F}_e$ and an inflation of $\mathcal{F}$ (more details about $\mathcal{F}_s$ are provided in the proof); clearly,  $\mathcal{F}_e$ tends to the set $\mathcal{F}$ with the decreasing of the error in the computation of the Jacobian and the measurement errors, which implies that $\mathcal{F}_s$ tends to $\mathcal{F}$ too. If these errors are small, the voltage violation is practically negligible.   

In the  case of no errors in the measurements and in the computation of the Jacobian, we have the following results. 

\begin{corollary}[\textit{Error-free implementation}]
\label{co:no_error}
    Let all the conditions in Theorem~\ref{thm:stability} be satisfied, and assume that there are no measurement errors, i.e., $E_M = 0$, and no errors in the Jacobian, i.e., $E_J = 0$ and. Let $\vecu(t)$, $t \geq t_0$,  be the unique trajectory of~\eqref{eq:controller}. Then it holds that
\begin{align*}
 \|\vecu(t)-\vecu^*\|&\leq  \sqrt{\frac{e_2}{e_1}} \|\vecu(t_0)-\vecu^*\|  e^{-  e_1 \eta s (t-t_0)} 
\end{align*}
and $\lim_{t \rightarrow + \infty } \|\vecu(t)-\vecu^*\| = 0$. \hfill $\triangle$
\end{corollary}

\begin{lemma}[\textit{Forward invariance in error-free implementation}]
\label{lem:forward_inv}
Let  the conditions in Theorem~\ref{thm:stability} be satisfied, and assume that there are no measurement errors (i.e., $E_M = 0$). Let $\vecu(t)$, $t \geq t_0$,  be the unique trajectory of~\eqref{eq:controller}, and $\vecV(t)$ be the corresponding trajectory of the voltages. Then,~\eqref{eq:controller} renders the set $\mathcal{F}$ forward-invariant. In particular: (i) if $\nu_i(t_0) \in [\underline{V}, \bar{V}]$, then $\nu_i(t) \in [\underline{V}, \bar{V}]$ for all $t \geq t_0$; (ii) if $\nu_i(t_0) \notin [\underline{V}, \bar{V}]$, then there exists $t^\prime \geq t_0$ such that $\nu_i(t) \in [\underline{V}, \bar{V}]$ for all $t \geq t^\prime$. \hfill  $\Box$
\end{lemma}

Corollary~\ref{co:no_error} quantifies the error in the convergence to $\vecu^*$, and certify local exponential stability properties for the proposed method. Lemma~\ref{lem:forward_inv} establishes that, for the case with no measurement errors and with the exact computation of the Jacobian matrix, the proposed method ensures that voltages are satisfied anytime.


\subsection{Extensions}
\label{sec:extensionVPP}

\emph{Virtual power plants}. In this section, we introduce an extension of our formulation for virtual power plants (VPPs). In this case, the goal is to coordinate the operation of the DERs to regulate voltages and to provide ancillary services to the bulk power system. In particular, the coordination is to ensure that the active and reactive powers at the substation track a reference $\{P_{0,set},Q_{0,set}\}$. The reference setpoint can be sent by the transmission system operator in order to provide frequency regulation or ancillary services. 

Using an expression for the powers at the substation such as $p_0 =  G_p(\vecu; \vecpl,\vecql)$ and $q_0 =  G_q(\vecu; \vecpl,\vecql)$, the OPF problem in \eqref{eq:OPFProb2} can be extended to include constraints of the form  $| (G_p(\vecu; \vecpl,\vecql) - P_{0,set})| \leq E_p$ and $|(G_q(\vecu; \vecpl,\vecql) - Q_{0,set}) |\leq E_q$, 
where $E_p>0$ and $E_q>0$ are tolerable  tracking errors for the setpoints $P_{0,set}$, $Q_{0,set}$, respectively. As an additional example, one can consider the constraint:
\begin{align}
        \left\| \left[
        \begin{array}{c}
            G_p(\vecu; \vecpl,\vecql)  \\
             G_q(\vecu; \vecpl,\vecql) 
        \end{array}
        \right] - 
        \left[
        \begin{array}{c}
            P_{0,set}  \\
            Q_{0,set} 
        \end{array}
        \right] 
        \right\| \leq E
\end{align}
with $E > 0$ again a given tracking error. In the proposed measurement-based SGF, the maps $G_p(\vecu; \vecpl,\vecql)$ and $G_q(\vecu; \vecpl,\vecql)$ would be replaced by measurements of the active and reactive powers at the substation, respectively. Moreover, the Jacobian matrix of $G_p(\vecu; \vecpl,\vecql)$ and $G_q(\vecu; \vecpl,\vecql)$ with respect to $\vecu$ can be approximated by using a linear model~\cite{LinearApprox2}. 

\emph{Other constraints}. The AC OPF formulation~\eqref{eq:OPFProb2} can be extended to include constraints on the power flows or currents. The proposed feedback-based   safe gradient flow can be naturally modified to enforce constraints on power flows or currents; to this end, one can use a linear model approximating the relationship between power injections at the DERs' nodes and power flows and currents in the controller design~\cite{bernstein2019real}. During the operational phase, the controller would rely on measurements of power flows and currents (or pseudo-measurements) in addition to voltages; see, for example, the discussion in~\cite{bernstein2019real}.

\section{Numerical experiments}
\label{sec:numericalexperiments}

\begin{figure*}[t!]
     \begin{subfigure}[b]{0.31\textwidth}
         \centering
         \includegraphics[width=1.0\linewidth]{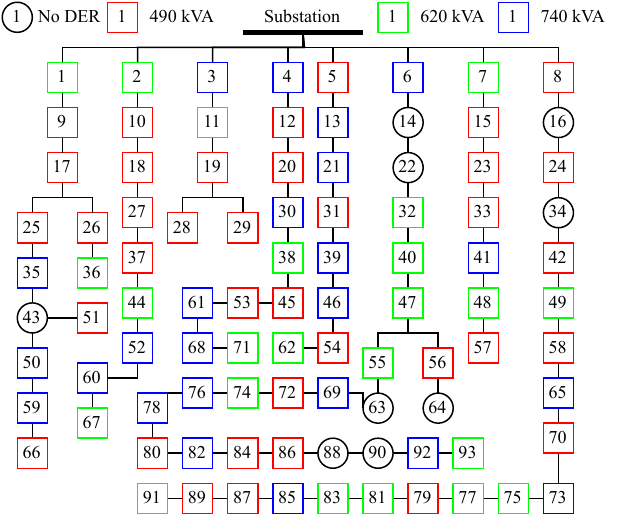}
         \caption{}
         \label{fig:network}
     \end{subfigure}
     \hfill
     \begin{subfigure}[b]{0.31\textwidth}
         \centering
         \includegraphics[width=1.0\linewidth]{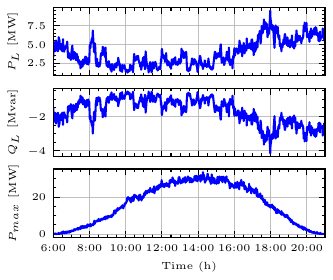}
         \caption{}
         \label{fig:Profiles}
     \end{subfigure}
     \hfill
     \begin{subfigure}[b]{0.31\textwidth}
         \centering
         \includegraphics{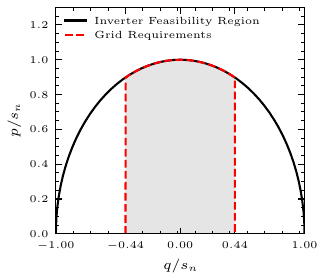}
         \caption{}
         \label{fig:PQ}
     \end{subfigure}
        \caption{(a) Distribution network used in the simulations. (b) Aggregated load consumption ($P_L,Q_L$) and PV production profiles ($P_{max}$) used in the simulations. (c) Operational set compared to grid code requirements inspired from the IEEE Std 1547-2018, where $s_n$ is the inverter rated power.}
        \label{fig:sim_setup}
\end{figure*}

We consider the medium voltage network (20 kV) shown in Fig.~\ref{fig:network}. We used a modified network from \cite{SimBench}, in which photovoltaic power (PV) plants have been randomly placed, with inverter-rated size picked randomly among $\{490,620,740\}$ kVA. The dynamics of the output power for the inverters are not implemented, as they are much faster than the controller dynamics; see, e.g.,~\cite{eggli2020stability}. Accordingly, when the controller updates the power setpoints, the inverter  implements them instantaneously. In the numerical experiments, we consider a system with PV plants; however, we note that  any type of inverter-interfaced DERs can be considered. Fig.~\ref{fig:Profiles} shows the aggregated loads and maximum available active power for PV plants throughout the day. The data is from the Open Power System 
Data\footnote{Data available at \url{https://data.open-power-system-data.org/household_data/2020-04-15}}, and have been modified to match the initial loads and PV plants nominal values present in the network. The reactive power demand is set such that the power factor is 0.9 (lagging). This would represent a typical summer day, with high PV production. We will show that, under these conditions, the electrical distribution network would undergo overvoltages.



\subsection{Simulation setup}

We compare the proposed measurement-based SGF with: (i) no control (NC); (ii) the online primal-dual method (PDM) proposed in \cite{dall2016optimal}; and, (iii) a Volt/Var control (VVC). We also compute the solution of a batch optimization (BO) method, where the AC OPF problem, with the power flow equations modeled using the nonlinear branch flow model \cite{BFM}, is solved.

\paragraph{Simulation parameters} The voltage service limits $\bar{V}$ and $\underline{V}$ are set to 1.05 and 0.95 p.u., respectively. The load and PV production profiles have a granularity of 10 seconds, i.e., active/reactive power consumption and maximum available active power for PV plants change every 10 seconds. For the SGF, it means that every 10 seconds, we pursue a new optimal solution. The SGF (using a forward Euler discretization), PDM and VVC algorithms are run every second. 

Based on the IEEE standard \emph{IEEE Std 1547-2018}, we consider the feasible set for the PV plants shown in Fig.~\ref{fig:PQ}. Although the inverter feasible set consists of a semicircle, there is no interest for PV owners to operate the PV plant at low power factors, i.e., large reactive power absorption/consumption and low active power production. Usually, PV plants are operated at unity power factor, i.e., on the vertical line passing through 0. The distribution system operator (DSO) often imposes grid requirements when a PV plant is connected to its network in order to provide support if needed. The grid requirements vary from one DSO to another. In this paper, we consider that the maximum reactive power the inverter can produce/consume is set to 44\% of its nominal apparent power. 
The vector-valued function modelling power limits is therefore
\begin{equation}
\ell_i(p_i,q_i) = \left[
\begin{array}{c}
     p_i^2+q_i^2 - s_{n,i}^2 \\
     p_i - p_{\text{max},i} \\ 
     -p_i \\
     -0.44s_{n,i}-q_i \\
     q_i-0.44s_{n,i} 
     \end{array} \right] . 
\end{equation}
It is assumed that $p_{\text{max},i}$ is known at the DERs. For example, one can use the method proposed in~\cite{hussain2018parameter} to estimate the maximum power point of PV arrays, and therefore, the maximum available power $p_{\text{max},i}$.
Finally, we consider the following cost function for the SGF, PDM, and BO:
\begin{equation}
    C_P(p_i,q_i) = \sum_{i\in\calG} c_p\left(\frac{s_{n,i}-p_i}{s_{n,i}}\right)^2 + c_q\left(\frac{q_i}{s_{n,i}}\right)^2
    \label{eq:costfunction}
\end{equation}
with $c_p = 3$ and $c_q = 1$. This cost function seeks to minimize active power curtailment and inverter power losses. The first part minimizes the active power curtailment, the second part the reactive power usage, which is also related to the inverter losses as less reactive power usage means less currents and thus, less Joules losses.

\begin{figure}[h!]
    \centering
    \includegraphics[width = 0.35\textwidth]{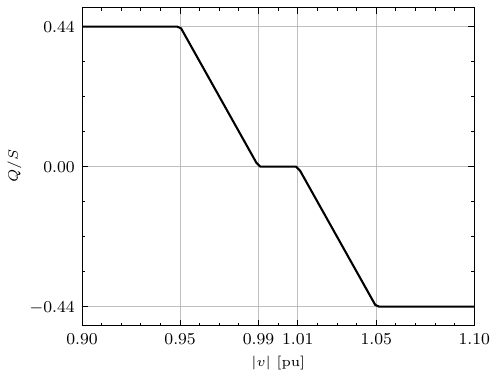}
    \caption{Implementation of IEEE standard \emph{IEEE Std 1547-2018} with $Q$ the reactive power injection, $S$ the nominal apparent power of the DER, $Q/S$ represents the ratio between $Q$ and $S$, and $|v|$ is the voltage magnitude at the node.}
    \label{fig:StaticVoltVar}
\end{figure}

\paragraph{Volt/Var control}
The Volt/Var control is inspired by the IEEE standard \emph{IEEE Std 1547-2018}. The parameters of the Volt/Var control have been adapted to match the voltage service limits considered in this paper. The maximum reactive power consumed/absorbed is set to 44\% of the nominal apparent power of the PV plant. The maximum power absorbed/produced is reached for voltages 1.05/0.95 pu, respectively. Finally, we implemented a deadband for voltages between 0.99 and 1.01 pu. 
Our implementation of the IEEE standard \emph{IEEE Std 1547-2018} is shown in Fig.~\ref{fig:StaticVoltVar}.

\paragraph{No control} For the no-control test case, we consider an overvoltage protection of PV plants, i.e., the plant is disconnected if the voltage level is too high. We consider three different status for the PV plant: \textit{running}, \textit{idling}, and \textit{disconnected}. When the PV plant is in status \textit{idling} or \textit{disconnected}, it does not inject active power or provide reactive power compensation. The disconnection scheme is inspired from the CENELEC EN50549-2 standard~\cite{cenelecstandard}, and has been adapted considering the voltage service limits used in this paper. The PV plant changes status from \textit{running} to \textit{disconnected} if: (i) the voltage at the point of connection goes above 1.06 pu, (ii) the root mean square value of the voltages measured at the point of connection for the past 10 minutes goes above 1.05 pu (the voltages are measured every ten seconds). 

The PV plant switches to status \textit{idling} if the voltage at the point of connection stays below 1.05 pu for 1 minute. To switch back to \textit{running} status, the PV plant has to be in \textit{idling} status. The switching to \textit{running} status occurs randomly in the  interval $[1 \text{min},10 \text{min}]$ (random, uniformly distributed). 

\begin{figure*}
    \centering
    \begin{subfigure}[b]{0.31\textwidth}
        \centering\includegraphics[width=1.1\linewidth]{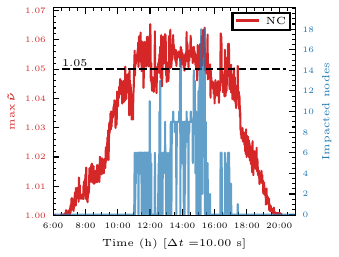}
        \caption{}
        \label{fig:NCov}
    \end{subfigure}
    \hfill
    \begin{subfigure}[b]{0.31\textwidth}
        \centering\includegraphics[width=1.1\linewidth]{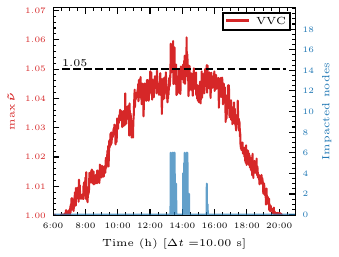}
        \caption{}
        \label{fig:VVCov}
    \end{subfigure}
    \hfill
    \begin{subfigure}[b]{0.31\textwidth}
        \centering\includegraphics[width=1.1\linewidth]{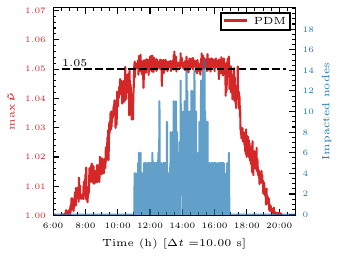}
        \caption{}
        \label{fig:PDMov}
    \end{subfigure}    

    \medskip

    \begin{subfigure}[b]{0.31\textwidth}
        \centering\includegraphics[width=1.1\linewidth]{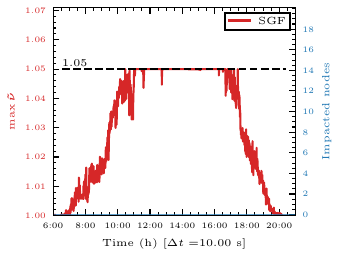}
        \caption{}
        \label{fig:SGFov}
    \end{subfigure}
    \hfill
    \begin{subfigure}[b]{0.31\textwidth}
        \centering\includegraphics[width=1.02\linewidth]{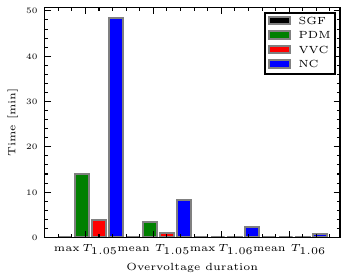}
        \caption{}        
        \label{fig:ovdur}
    \end{subfigure}
    \hfill
    \begin{subfigure}[b]{0.31\textwidth}
    \centering
    \includegraphics[width=1.02\linewidth]{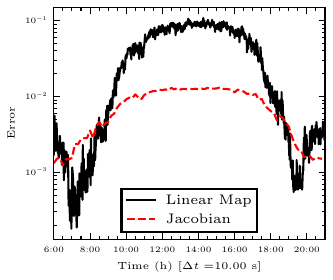}
    \caption{}
    \label{fig:Maperror}
    \end{subfigure}
    \caption{(a) Overvoltages for NC. (b) Overvoltages for VVC. (c) Overvoltages for PDM. (d) Overvoltages for SGF. (e) Overvoltage duration times. (f) Linear map error: $||\hat{H}(\vecu;\vecpl,\vecql) - H(\vecu;\vecpl,\vecql)||$ and Jacobian error: $||J_{\hat{H}} - J_{H}(\vecu;\vecpl,\vecql)||$, where  $\vecu$ is picked from the SGF algorithm.}
    \label{fig:OV}
\end{figure*}

\subsection{Results}
In the following, we compare the different methods in terms of their cost function values, the system losses, and the voltage levels. 
Notice that system losses are not integrated in the cost function~\eqref{eq:costfunction}, since they conflict with the term that considers active power curtailment. In this paper, we design the cost function to promote renewable energy resources, hence maximizing the solar production while keeping voltage levels within pre-defined bounds. However, the DSO is also concerned by the system losses. Thus, one needs to look how the different methods perform with respect to them.

\paragraph{Voltage regulation} In Fig.~\ref{fig:OV}, one can see the maximum voltage observed at every time step, as well as the number of impacted nodes where we observed a voltage greater than $\bar{V}$. It can be seen that only the SGF method does not lead to voltage violations. This is precisely because our approach is based on the theory of control barrier function. PDM leads to voltage violations not only because of the transient of the dual variables, but also because it is designed based on a regularization of the Lagrangian function as explained in~\cite{dall2016optimal}. VVC performs well, although as shown hereafter, it leads to larger system losses and a greater cumulative cost than SGF or PDM. The overall voltage profile is also shifted downward due to its proportional feedback control. One can observe the spikes in the NC method due to multiple disconnections of DERs because of a prolonged overvoltage duration. Finally, one can see that with PDM, the voltages oscillate around the threshold voltage of 1.05 pu. 

\paragraph{Over-voltage duration} In Fig.~\ref{fig:ovdur}, we show the duration of overvoltages. We define $T_{\geq \alpha; i}$ as a vector containing the number of consequent time steps during which node $i$ sees his voltage above the value $\alpha$. The value $\max T_{\geq \alpha}$ corresponds to the maximum value among all $T_{\geq \alpha; i}$ for $i\in \mathcal{N}$ and corresponds to the maximum consequent time duration during which one nodal voltage was above $\alpha$.  The value $\text{mean}\ T_{\geq \alpha}$ is the maximum of the mean absolute values of every vector $T_{\geq \alpha; i}$ for $i\in \mathcal{N}$, representing the average time duration of overvoltage. Since SGF algorithm does not yield overvoltages, it does not appear on this graph. One can see that the NC method does not perform well, as the active power curtailment is activated only for large overvoltage (above 1.06pu) or for prolonged overvoltage (above 1.05pu). 

\begin{figure}[h!]
    \centering
    \includegraphics[width=.8\linewidth]{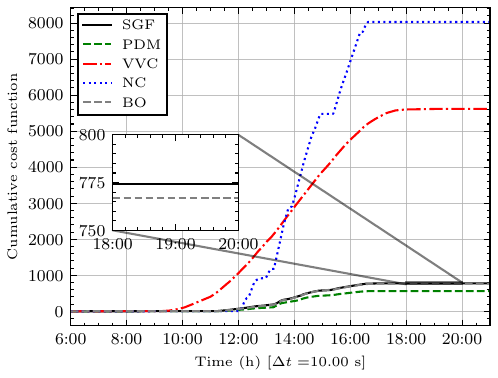}
    \caption{Achieved values of cumulative cost~\eqref{eq:costfunction}.}
    \label{fig:cost}
\end{figure}

\paragraph{Achieved cost} We show the cumulative cost function in Fig.~\ref{fig:cost}, i.e., the cumulative sum of the cost function at every time step. It is clear that the NC method leads to the largest cumulative costs, as its implementation leads to full curtailment of solar production and no usage of reactive power. The VVC shows the second highest cost because of its inefficient usage of reactive power reserves. We have to bear in mind that these two solutions cannot practically achieve the optimal solutions of the BO method since they can only play with either the active or reactive power output of solar inverters. Furthermore, they are, by design, decentralized control algorithms, and  do not have full information of the system state. We observe that PDM has the lowest cumulative cost, which comes at the detriment of voltage violations, as observed in Fig.~\ref{fig:PDMov}. The SGF cumulative cost superposes the BO cost.

\begin{figure}[h!]
    \centering
    \includegraphics[width=.8\linewidth]{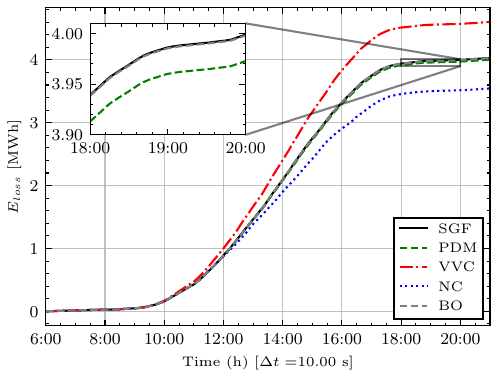}
    \caption{Sum of energy losses throughout the day}
    \label{fig:Loss}
\end{figure}

\paragraph{System losses}
The cumulative system losses for the different methods are shown in Fig.~\ref{fig:Loss}. The NC method leads to the lowest system losses as it drastically reduces the amount of active power flows in the network by fully curtailing solar production. The VVC leads to the highest system losses as it over-uses reactive power compensation to mitigate voltage issues. This results in larger power flows throughout the network, hence larger power system losses. We can observe that PDM, SGF, and BO have similar system losses.

Finally, Fig.~\ref{fig:Maperror} shows the error between the linear approximation of the power flow equations and the non-linear power flow equations, validating our choice for the linear map. It also shows the error between the approximate Jacobian (which is constant), and the true Jacobian computed numerically.

\section{Conclusions}
\label{sec:conclusion}

This paper has addressed the problem of continuously adjusting the power outputs of DERs to pursue feasible  solutions of AC OPF problems. We have employed a continuous approximation of projected gradient flows, modified to accommodate voltage measurements from the electrical network, to ensure the satisfaction of voltage constraints at all times.  We showed practical exponential stability for scenarios where voltage measurements are subject to errors, and where only an approximation of the Jacobian matrix of the power flow equations is available. Our method was experimentally validated on a 93-bus distribution system with realistic load and production profiles. Our approach exhibited a performance significantly superior in terms of voltage regulation to existing online primal-dual methods and Volt/Var strategies. Future research efforts will look at data-driven implementations and event-triggered implementations of the feedback-based safe gradient flow. 


\appendix

\subsection{CBF-based design principles}
\label{sec:design}

In this section, we provide insights on the CBF-based design approach for the proposed measurement-based SGF. Consider rewriting the OPF problem defined in \eqref{eq:OPFProb2} in the following general form:
\begin{equation}\label{eq:NLopt}
    \begin{aligned}
        \min_{x\in \real^{2G}}\quad & f(x)\\
        \textrm{s.t.} \quad & g(x) \leq 0
    \end{aligned}
\end{equation}
with $f : \real^{2G} \rightarrow \real$ and $g : \real^{2G} \rightarrow \real^{P}$, where $P$ is the number of voltage and power constraints. Let $\mathcal{F} = \{x \in \real^{2 G}\ |\ g(x) \leq 0\}$ and  $x^*$ be a local optimizer of \eqref{eq:NLopt}. This point, along with the optimal dual variables $y^* \in \real^{P}$, satisfy the  \textit{Karush-Kuhn-Tucker} conditions:
\begin{equation}\label{eq:KKT}
    \begin{aligned}
        \nabla f(x^*) + \frac{\partial g(x^*)^\top}{\partial x} y^* = 0\\
        g(x^*) \leq 0 \\
        y^*\geq 0, \,\, (y^*)^\top g(x^*) = 0\\
    \end{aligned}
\end{equation}
As proposed in~\cite{allibhoy2023control}, the optimization problem \eqref{eq:NLopt} can be solved using nonlinear dynamics of the form:
\begin{equation}\label{eq:ConceptSGF}
    \begin{aligned}
        \dot{x} = -\nabla f(x) - \frac{\partial g(x)^\top}{\partial x} y
    \end{aligned}
\end{equation}
which  can be interpreted as a modification of the  gradient flow $-\nabla f(x)$, where the input  $y$ can be designed to ensure that the set $\mathcal{F}$ is forward invariant. To this end,  define the following admissible set for $y$:
\begin{equation}
        K_{\beta}(x) := \Big\{y \in \real^{P}_{\geq 0} \bigg\rvert -\frac{\partial g}{\partial x}\frac{\partial g}{\partial x}^\top y \leq \frac{\partial g}{\partial x} \nabla f(x) - \beta g(x)\Big\} ,
\end{equation}
where $\beta>0$ is a design parameter, which is inspired by CBF arguments~\cite{ames2019control}; see~\cite{allibhoy2023control}. Since we want  the drift term $\frac{\partial g(x)^\top}{\partial x} y$ as small as possible while ensuring that the set $\mathcal{F}$ is feasible, the input is computed as~\cite{allibhoy2023control}:
\begin{equation}\label{eq:Controller}
        y(x) = \arg \min_{y\in K_{\beta}(x)} \left\|\frac{\partial g(x)^\top}{\partial x} y\right\|^2
\end{equation}
for each $x$. The overall modified gradient flow is then given by \eqref{eq:ConceptSGF} with the input $y(x)$ in \eqref{eq:Controller}.


In \cite{allibhoy2023control}, it is  shown that 
\eqref{eq:ConceptSGF} with the input $y(x)$ in \eqref{eq:Controller} is equivalent to dynamics of the form $\dot x = F_{\beta} (x)$, where the flow $F_{\beta}(x)$ is defined as: 
\begin{equation}\label{eq:SGFimpl}
    \begin{aligned}
        F_{\beta}(x) := \arg \min_{\theta \in \real^{2G}}\quad \frac{1}{2}\left\|\theta + \nabla f(x)\right\|^2\\
        \textrm{s.t.}\quad \frac{\partial g(x)}{\partial x}^\top\theta \leq -\beta g(x)
    \end{aligned}
\end{equation}
In this paper, we leverage dynamics of the form~\eqref{eq:SGFimpl} to solve our AC OPF problem; however, as explained in Section~\ref{sec:safeopfpursuit}, the dynamics are modified to accommodate measurements.

\subsection{Proof of Theorem~\ref{thm:stability}}
\label{ap:proof}

We recall that $\vecV$ is a short-hand notation for the real voltages, i.e., $\vecV = H(\vecu;\vecpl,\vecql)$, and $\Tilde{\vecV}$ is the vector of (pseudo-)measurements. Recall also that $F_\beta(\vecu,\vecV) = \bar{F} _\beta(\vecu,\zero,\zero)$ and $\hat{F}_{\beta}(\vecu, \tilde \vecV) = \bar{F}_\beta(\vecu,J_{\hat{H}}(\vecu)\!-\! J_H(\vecu),\tilde \vecV\!-\!\vecV)$. First, we express our controller as:
\begin{align*}
  \dot \vecu & = \eta \hat{F}_\beta(\vecu,\tilde \vecV) \\
  & = \eta \bar{F} _\beta(\vecu,\zero,\zero) \\
  & \quad + \eta[\bar{F}_\beta(\vecu,J_{\hat{H}}(\vecu)\!-\! J_H(\vecu),\tilde \vecV\!-\!\vecV) - \bar{F}_\beta(\vecu,\zero,\tilde \vecV\!-\!\vecV)] \\
  & \quad + \eta[\bar{F}_\beta(\vecu,\zero,\tilde \vecV\!-\!\vecV) - \bar{F} _\beta(\vecu,\zero,\zero)] 
\end{align*}
where we added and subtracted $\bar{F} _\beta(\vecu,\zero,\zero)$ and $\bar{F}_\beta(\vecu,\zero,\tilde \vecV\!-\!\vecV)$, and we re-organized the terms. The feedback-based SGF can then be understood as a perturbation of the nominal gradient flow $\bar{F} _\beta(\vecu,\zero,\zero)$.

By \cite[Lemma 5.11 and Theorem 5.6(iii)]{allibhoy2023control}, $\bar{F} _\beta(\vecu,\zero,\zero) $ is differentiable at $\vecu^*$ and its Jacobian $E=\frac{\partial \bar{F} _\beta(\vecu,\zero,\zero) }{\partial \vecu}\mid_{\vecu=\vecu^*}$ is negative definite. 
Recall that $e_1 =-\lambda_{\max}(E)$ and $e_2 =-\lambda_{\min}(E)$. Let $P:=\int_0^\infty (\exp(E\zeta)^\top  \exp(E\zeta) d\zeta$, and then by \cite[Theorem 4.12]{khalil2002nonlinear}, it holds that $ PE+E^\top P=-\mathbf{I}_n   $, and $
   \frac{1}{2e_2} \|u-u^*\|_2^2 \leq (u-u^*)^\top P (u-u^*)\leq \frac{1}{2e_1} \|u-u^*\|_2^2 $.
Let $V_1(\vecu):=$$(\vecu-\vecu^*)^\top P (\vecu-\vecu^*)$; then we bound $2 (\vecu-\vecu^*)^\top P \bar{F} _\beta(\vecu,\zero,\zero) $ and then leverage this bound to estimate $ \dot{V}_1$:
\begin{align*}
    &\hspace{-.5cm} 2 (\vecu-\vecu^*)^\top P \bar{F} _\beta(\vecu,\zero,\zero)\\
    &= (\vecu - \vecu^*)^\top \left( P E + E^\top P \right) (\vecu - \vecu^*) \\
    & \quad  + 2 (\vecu - \vecu^*)^\top P \hat{g}(\vecu)\\
    &\leq  - \|\vecu - \vecu^*\|^2 + \frac{1}{e_1}\|\vecu-\vecu^*\| L \|\vecu - \vecu^*\|^2\\
     & \leq  \left(-1+ \frac{ L}{e_1}\left\|\vecu-\vecu^*\right\|_2\right)\left\|\vecu-\vecu^*\right\|_2^2 \leq  -s  \|\vecu - \vecu^*\|^2 
\end{align*}
where the last inequality holds  if $\|\vecu-\vecu^*\|\leq \frac{e_1}{L}(1-s)$, for any $s\in(s_{\min},1]$. Then, 
\begin{align*}
    \dot{V}_1&=2 (\vecu-\vecu^*)^\top P \dot u\\
    &= 2\eta (\vecu-\vecu^*)^\top P  \bar{F} _\beta(\vecu,\zero,\zero)) \\
    & + 
   2\eta (\vecu-\vecu^*)^\top P  [\bar{F}_\beta(\vecu,J_{\hat{H}}\!-\! J_H,\tilde \vecV\!-\!\vecV) - \bar{F}_\beta(\vecu,\zero,\tilde \vecV\!-\!\vecV)]\\ 
    & +2\eta (\vecu-\vecu^*)^\top P  
    [\bar{F}_\beta(\vecu,\zero,\tilde \vecV\!-\!\vecV) - \bar{F} _\beta(\vecu,\zero,\zero)]\\
    & \leq - \eta s\|\vecu-\vecu^*\|^2 + 2\eta \ell_{F_J} \|\vecu-\vecu^*\| \| P\| \|J_H-J_{\hat{H}} \|\\
    & + 2\eta \ell_{F_v} \|\vecu-\vecu^*\| \| P\| \|\tilde{\vecV}-\vecV \| \\
    &\leq  -\eta s\|\vecu-\vecu^*\|^2+\eta\frac{\ell_{F_J}E_J+\ell_{F_v}E_M   }{e_1}\|\vecu-\vecu^*\|\\
    &\leq  -2e_1\eta s V_1+\eta\sqrt{2e_2}\frac{\ell_{F_J}E_J+\ell_{F_v}E_M   }{e_1}\sqrt{V_1} . 
\end{align*}
Define $V_2(\vecu):=\sqrt{V_1(\vecu)}$. Then,  
\begin{align*}
\dot{V}_2=\frac{\dot{V}_1 }{2\sqrt{V_1}}\leq& \frac{-2e_1\eta s V_1+\eta\sqrt{2e_2}\frac{\ell_{F_J}E_J+\ell_{F_v}E_M   }{e_1}\sqrt{V_1} }{2\sqrt{V_1}}\\
=&-e_1\eta s V_2+\eta\sqrt{2e_2}\frac{\ell_{F_J}E_J+\ell_{F_v}E_M   }{2e_1}
\end{align*}

In addition, we note that that for any $a\geq 0,~b> 0$,  $y(t)=y(t_0)\exp{(-b(t-t_0) )}+\frac{a}{b}(1-\exp{(-b(t-t_0) })$ is the solution of $\dot y=-by+a, y(t_0)=y(t_0)$. Hence by the  
Comparison Lemma \cite[Lemma 3.4]{khalil2002nonlinear}, it follows that 
\begin{align*}
V_2(t) & \leq V_2(t_0) e^{- e_1\eta s(t-t_0)} \\
& \quad +\frac{\sqrt{2e_2}(\ell_{F_J}E_J+\ell_{F_v}E_M   )}{2 s e_1^2} \left(1-e^{- e_1\eta s(t-t_0)}     \right).
\end{align*}

Thus, one has that
\begin{align*}
 & \hspace{-1.0cm} \|\vecu(t)-\vecu^*\| \leq \sqrt{2e_2} V_2(t)\\
 \leq & \sqrt{2e_2} V_2(t_0) e^{- e_1\eta s(t-t_0)}\\
 &+  \frac{2e_2(\ell_{F_J}E_J+\ell_{F_v}E_M   )}{2 s e_1^2} \left(1-e^{- e_1\eta s(t-t_0)}    \right ) \\
\leq & \sqrt{\frac{1}{2e_1}}\sqrt{2e_2}  e^{- e_1\eta s(t-t_0)}\|\vecu(t_0)-\vecu^*\|\\
 &+ \frac{e_2(\ell_{F_J}E_J+\ell_{F_v}E_M   )}{ s e_1^2} \left(1-e^{- e_1\eta s(t-t_0)}     \right)\\
 =& \sqrt{\frac{e_2}{e_1}} e^{- e_1\eta s(t-t_0)}\|\vecu(t_0)-\vecu^*\|\\
 &+ \frac{e_2(\ell_{F_J}E_J+\ell_{F_v}E_M   )}{ s e_1^2} \left(1-e^{- e_1\eta s(t-t_0)}     \right) .
\end{align*}
which proves the result. The limits for $t\to+\infty$ can be computed straightforwardly.  

\subsection{Proof of Lemma~\ref{lem:pratical_forward}}
\label{ap:pratical_forward}

The proof leverages Nagumo's Theorem~\cite{nagumo1942lage}. 
For the feedback-based SGF $\hat{F}_\beta(\vecu,\vecV)$ in~\eqref{eq:controller_approx}, it holds that $-\nabla \hat{H}_i(\vecu)^\top\hat{F}_\beta \leq -\beta \left(\underline{V}-\tilde \nu_i\right)$.   Recall that $\tilde \nu_i= H(\vecu;\vecpl,\vecql) + e_i$, for $i\in\mathcal{M} $. It follows that
\begin{align*}
    & -\nabla \hat{H}_i(\vecu)^\top\hat{F}_\beta \leq -\beta \left(\underline{V}-{H}(\vecu;\vecpl,\vecql)-e_i\right) \\
    &= -\beta (\underline{V}-\hat{H}(\vecu;\vecpl,\vecql)-e_i +({H}(\vecu;\vecpl,\vecql)-\hat{H}(\vecu;\vecpl,\vecql)))\\
    &\leq -\beta ( (\underline{V}-E_M-E_{\hat{H}})-\hat{H}(\vecu;\vecpl,\vecql))
\end{align*}
where $E_{\hat{H}}:=\max_{\vecu\in \mathcal{U}} \| {H}(\vecu;\vecpl,\vecql)-\hat{H}(\vecu;\vecpl,\vecql)\|$. Similarly, it also holds that $\nabla \hat{H}_i(\vecu)^\top\hat{F}_\beta \leq -\beta (\hat{H}(\vecu;\vecpl,\vecql)- (\bar{V}+E_M+E_{\hat{H}}))$. 
Thus, the set 
\begin{align*}
\mathcal{F}_s:= &\{\vecu: \underline{V}-E_M-E_{\hat{H}}\leq \hat{H}_i(\vecu;\vecpl,\vecql) \leq \bar{V}+E_M+E_{\hat{H}},
\\
 & \forall i \in \mathcal{M}, \vecu \in \mathcal{U}\}
\end{align*}
is forward invariant under~\eqref{eq:controller_approx}. Note that $ \mathcal{F}_s$ is a subset of~$\mathcal{F}_e$, and this concludes the proof.

\bibliographystyle{IEEEtran}
\bibliography{references}

\end{document}